\documentclass[journal,10pt]{IEEEtran}
\usepackage{multicol}
\usepackage{multirow}
\usepackage{array}
\usepackage{amsmath}
\usepackage{algorithm}
\usepackage{algpseudocode}
\usepackage{amssymb}
\usepackage{amsfonts}
\usepackage{textcomp}
\usepackage{xcolor}
\usepackage{stfloats}
\usepackage{url}
\usepackage{verbatim}
\usepackage{graphicx}
\usepackage{caption}
\usepackage{subcaption}
\def\BibTeX{{\rm B\kern-.05em{\sc i\kern-.025em b}\kern-.08em
    T\kern-.1667em\lower.7ex\hbox{E}\kern-.125emX}}
\usepackage{balance}

\begin{document}

\title{Bridging 6G IoT and AI: LLM-Based Efficient Approach for Physical Layer's Optimization Tasks}
\author{
{Ahsan~Mehmood}, {Naveed Ul Hassan, {\em Senior Member, IEEE}}, {Ghassan M. Kraidy, {\em Senior Member, IEEE}}

\thanks{(Corresponding Author: Ahsan Mehmood)

Ahsan Mehmood and Ghassan M. Kraidy are with the Department of Electronic Systems, Norwegian University of Science and Technology, Norway.
(e-mail:\{ahsan.mehmood, ghassan.kraidy\}@ntnu.no) 

Naveed Ul Hassan is with the Department of Electrical Engineering, Lahore University of Management sciences (LUMS), Pakistan (emails: {naveed.hassan}@lums.edu.pk.) }

}

 

\maketitle

\begin{abstract}

This paper investigates the role of large language models (LLMs) in sixth-generation (6G) Internet of Things (IoT) networks and proposes a prompt-engineering-based real-time feedback and verification (PE-RTFV) framework that  perform physical-layer's optimization tasks through an iteratively process. By leveraging the naturally available closed-loop feedback inherent in wireless communication systems, PE-RTFV enables real-time physical-layer optimization without requiring model retraining. The proposed framework employs an optimization LLM (O-LLM) to generate task-specific structured prompts, which are provided to an agent LLM (A-LLM) to produce task-specific solutions. Utilizing real-time system feedback, the O-LLM iteratively refines the prompts to guide the A-LLM toward improved solutions in a gradient-descent-like optimization process. We test PE-RTFV approach on wireless-powered IoT testbed case study on user-goal-driven constellation design through semantically solving rate-energy (RE)-region optimization problem which demonstrates that PE-RTFV achieves near–genetic-algorithm performance within only a few iterations, validating its effectiveness for complex physical-layer optimization tasks in resource-constrained IoT networks.




\end{abstract}

\section{INTRODUCTION}
\label{sec:introduction}
The remarkable success of transformer models has driven the rapid advancement of large language models (LLMs), which are capable of extracting knowledge from vast datasets and demonstrating exceptional reasoning abilities. Models such as GPT-2/3/4, Falcon LLM, LLaMA (Large Language Model Meta AI), and visual models like DALL·E have enabled a wide range of novel applications \cite{11152695}. Building on these developments, multi-modal LLMs have also emerged that can process diverse input modalities including text, audio, video, and sensory data. 

As a driving force behind future innovation, LLMs are expected to play a pivotal role in reshaping wireless network architectures and their functionalities. Specifically, multi-modal LLMs trained on datasets containing radio signals, communication standards, images, and audio are anticipated to support advanced functions such as proactive localization, holographic beamforming, efficient resource management and so on. Furthermore, generative artificial intelligence (GenAI) and LLM models are poised to drive operational, environmental, and computational intelligence within sixth generation (6G) systems \cite{10980288}. These models are expected to be deployed across a wide range of 6G devices and network components, enhancing robustness and enabling intelligence throughout the network. Ultimately, 6G networks will consist of intelligent agents that collaborate and adapt in real time, fostering collective intelligence \cite{10980288} and shifting communication systems from data- and model-centric architectures to knowledge- and reasoning-driven frameworks. 

Despite the remarkable success of LLMs across various domains, their application in 6G internet of things (IoT) networks remains largely unexplored. Research in this direction is only beginning to emerge as this is a new and evolving area. Several recent studies have presented surveys, visions, and research directions for leveraging LLMs in 6G IoT networks. The current body of work has only begun to uncover the broader potential of LLMs in 6G IoT communications. For example, a split federated learning strategy is introduced for fine-tuning LLMs within a 6G IoT framework in \cite{chen2025llm}. Similarly, \cite{10705427} presents an in-context learning (ICL)-based prompting method; however, the evaluation is limited to a table-based question answering task. Despite these initial contributions, substantial opportunities remain for further research. This is particularly true for physical-layer's optimization tasks which involve complex reasoning and require tailored adaptation strategies to fully leverage the capabilities of LLMs in 6G IoT networks.

Recent works have explored the use of general-purpose LLMs for optimization tasks. In \cite{lee2026llm}, an in-context learning (ICL) approach was applied to power allocation by providing channel gains and their corresponding optimal transmit powers as reference examples. However, this method requires prior access to optimal solutions, which may not be available for many practical optimization problems.
Another line of work \cite{11031194} proposed an LLM-based framework for rate and proportional fairness maximization in multi-user uplink systems, where the LLM generates candidate solutions that are refined using an internal objective-function evaluator. While effective in static settings, the reliance on a predefined evaluator limits adaptability to time-varying network conditions. 

More recently, \cite{peng2025llm} addressed non-convex optimization by converting textual problem descriptions into mathematical formulations and then into convex problems using techniques such as semidefinite relaxation (SDR) and successive convex approximation (SCA). The resulting problems are solved using LLM generated MATLAB or Python code, followed by validation and feasibility correction. Although powerful, this solver-centric pipeline incurs significant computational complexity, hindering real-time deployment.
In contrast, we propose a prompt-engineering-based real-time feedback and verification (PE-RTFV) framework that enables lightweight, iterative solution refinement using real-time feedback alone. Unlike existing approaches, PE-RTFV does not require prior optimal examples, internal objective evaluators, or explicit optimization solvers. We demonstrate its effectiveness through a user-goal-oriented constellation design problem in a 6G wireless-powered IoT (WP-IoT) network.


WP-IoT networks support heterogeneous devices, including energy-harvesting (EH) nodes, conventional information transceivers, and simultaneous wireless information and power transfer (SWIPT) devices. These devices harvest energy by passing the received signal through nonlinear EH rectifier circuits, whose operating regime depends on the input signal power level \cite{8815474}. At low input powers, rectifier diodes operate in the square-law and transition regions, whereas higher input powers drive the rectifier toward saturation, rendering the harvested energy highly sensitive to the waveform characteristics of the received signal. Combined with the heterogeneous energy-harvesting requirements across WP-IoT devices, this nonlinear behavior necessitates device-specific waveform shaping through modulation design. While standard modulation schemes such as M-QAM are well suited for conventional information transmission, they are generally unsuitable for SWIPT devices, which rely on a single radio-frequency signal for simultaneous information decoding and energy harvesting and therefore require SWIPT-optimized constellation designs. Consequently, constellation design and optimization for WP-IoT and SWIPT systems has attracted significant research interest.

Several early studies have addressed this problem by proposing asymmetric PSK (APSK) and asymmetric circular QAM constellations to enhance energy-harvesting performance \cite{bayguzina2019asymmetric, 9951152}, followed by full-phase asymmetric QAM \cite{8815474} and limited-phase asymmetric QAM designs \cite{mehmood2025asymmetric}. However, these approaches typically rely on accurate energy-harvesting models, predefined energy thresholds, and quality-of-service (QoS) constraints, which may be unavailable or time-varying in practical deployments. Moreover, the high computational complexity of existing constellation design algorithms (e.g., genetic algorithms) limits design flexibility and prevents real-time modulation adaptation to the diverse energy-harvesting requirements across devices. In contrast, the proposed PE-RTFV  enables lightweight and flexible optimization framework for optimizing constellation shape by learning directly from feedback, allowing real-time adaptation to heterogeneous energy-harvesting demands without requiring precise device models or computationally expensive optimization routines.
Specifically, the key contributions of this work are summarized below:
\begin{itemize}
    \item We review the transformative role of LLMs in enabling operational, environmental, and computational intelligence in 6G IoT networks, and also discusses the task-specific adaptation techniques such as parameter fine-tuning (PFT) and prompt engineering (PE).
      
    \item We then introduce the PE-RTFV framework for physical layer's resource allocation and optimization tasks, which exploits the inherent closed-loop feedback and control mechanisms of wireless communication systems.
    
    \item We demonstrate the effectiveness of the proposed PE-RTFV framework through detailed case studies on modulation design for SWIPT devices by performing rate energy (RE)-region optimization. For rigorous evaluation, we conduct independent experiments under three feedback scenarios: i) full feedback, ii) codebook-based feedback, and iii) quantized feedback.
    
    \item Finally, we benchmark the PE-RTFV framework against the state-of-the-art genetic algorithm (GA), demonstrating that the proposed approach can efficiently solve highly complex optimization tasks relying solely on feedback information.
\end{itemize}

The remainder of this paper is organized as follows. Section~\ref{sec2} discusses AI capabilities in 6G and the role of LLMs in 6G IoT systems. Section~\ref{sec3} presents task-specific model adaptation approaches. Section~\ref{sec4} details the proposed PE-RTFV framework. Section~\ref{sec:casestudy} provides a case study evaluating the performance of PE-RTFV for modulation design via RE-region optimization. Finally, Section~\ref{conc} concludes the paper.

\section{LLMs in 6G IoT Networks}
\label{sec2}
In 6G networks, LLMs can serve as intelligent control agents, enabling autonomous optimization across heterogeneous IoT services by interpreting system states and feedback. In the following, we outline their role in 6G IoT networks.

\subsection{Roles of LLMs}
6G IoT networks are expected to exhibit intelligence across operational, environmental, and computational domains. Operational intelligence governs core network functions such as resource allocation, beamforming, access control, constellation design, and user scheduling. Conventional approaches rely on offline optimization and fixed control policies, which perform well in static scenarios but lack adaptability in dynamic environments, while real-time re-optimization is often infeasible for resource-constrained IoT devices. In contrast, LLMs enable context-aware decision-making by leveraging historical information and evolving network conditions, supporting adaptive control of power, spectrum, and beamforming.

Beyond operational control, environmental intelligence allows networks to actively shape the wireless propagation environment. Technologies such as reconfigurable intelligent surfaces and aerial platforms enable dynamic manipulation of signal paths and network topology \cite{9424177}. When combined with LLM-driven control, these resources can be configured in real time to improve coverage, energy efficiency, and link reliability, fostering a tighter interaction between the network and its physical environment.

Computational intelligence in wireless systems has traditionally relied on heuristic and metaheuristic optimization, which often face scalability and adaptability limitations. LLMs offer a complementary paradigm by enabling reasoning-driven decision-making that can support or partially replace conventional algorithms. However, challenges related to mathematical accuracy, computational efficiency, and memory overhead remain \cite{ahn2024large}, and addressing these issues is critical for the practical deployment of LLM-based intelligence in 6G IoT networks.

\subsection{LLM Task-Specific Adaptation Techniques}\label{sec3}
LLMs exhibit strong generalization due to large-scale pretraining, and their adaptation to domain-specific data is achieved through PE and PFT, discussed in detail below. make it one line
\subsubsection{Parameter-Efficient Fine-Tuning }
This approach requires retraining a foundational LLM using task-specific data. The most direct method, full model fine-tuning (MFT), updates all trainable parameters but is impractical for distributed IoT networks due to the need to retrain and store billions of parameters per task, leading to prohibitive computational and storage overhead. To mitigate these limitations, several lightweight parameter-efficient fine-tuning (PFT) methods have been proposed \cite{zhao2023survey}. For instance, Low-Rank Adaptation (LoRA) introduces small trainable low-rank matrices into transformer layers while keeping pretrained weights frozen, significantly reducing training cost and memory usage \cite{hu2022lora}. Similarly, adapter tuning inserts compact task-specific modules into the transformer architecture and trains only these adapters \cite{hu2023llm}. Prefix tuning provides an even lighter alternative by optimizing a continuous task-specific prefix while freezing all model parameters, tuning as little as $0.01\%$ of the parameters compared to approximately $3.6\%$ for adapter-based methods.

\subsubsection{Prompt Engineering (PE)}
In prompt engineering (PE), a general-purpose pretrained LLM (e.g., ChatGPT) is guided to perform task-specific objectives solely through carefully designed prompts, without modifying model parameters. By eliminating retraining and the need to store multiple fine-tuned models for each task, PE is well suited for distributed IoT and 6G networks with dynamic tasks and operating conditions. Two widely used PE techniques are in-context learning (ICL) and chain-of-thought (CoT) prompting, which differ in how they steer model behavior.

ICL enables task adaptation by embedding a small set of task-relevant input–output examples in the prompt, allowing analogy-based generalization without parameter updates. CoT prompting, in contrast, guides the model through intermediate reasoning steps, improving performance on complex tasks but potentially producing inconsistent reasoning under uncertainty, which has motivated refinement techniques based on sampling and verification \cite{zhao2023survey}.  

Compared to PFT, which introduces additional trainable parameters and task-specific model instances, PE achieves task specialization without increasing model complexity or storage overhead. This makes PE particularly attractive for real-time physical-layer optimization problems in wireless systems—such as power allocation, beamforming, and constellation design—that require rapid, feedback-driven adaptation. Building on these advantages, we introduce in the next section a PE-based real-time feedback and verification (PE-RTFV) framework for efficient online optimization without retraining.

\section{PE-based Real-Time Feedback and Verification Approach}\label{sec4}
\begin{figure}[htb]
\centering
\includegraphics[width=0.5\textwidth]{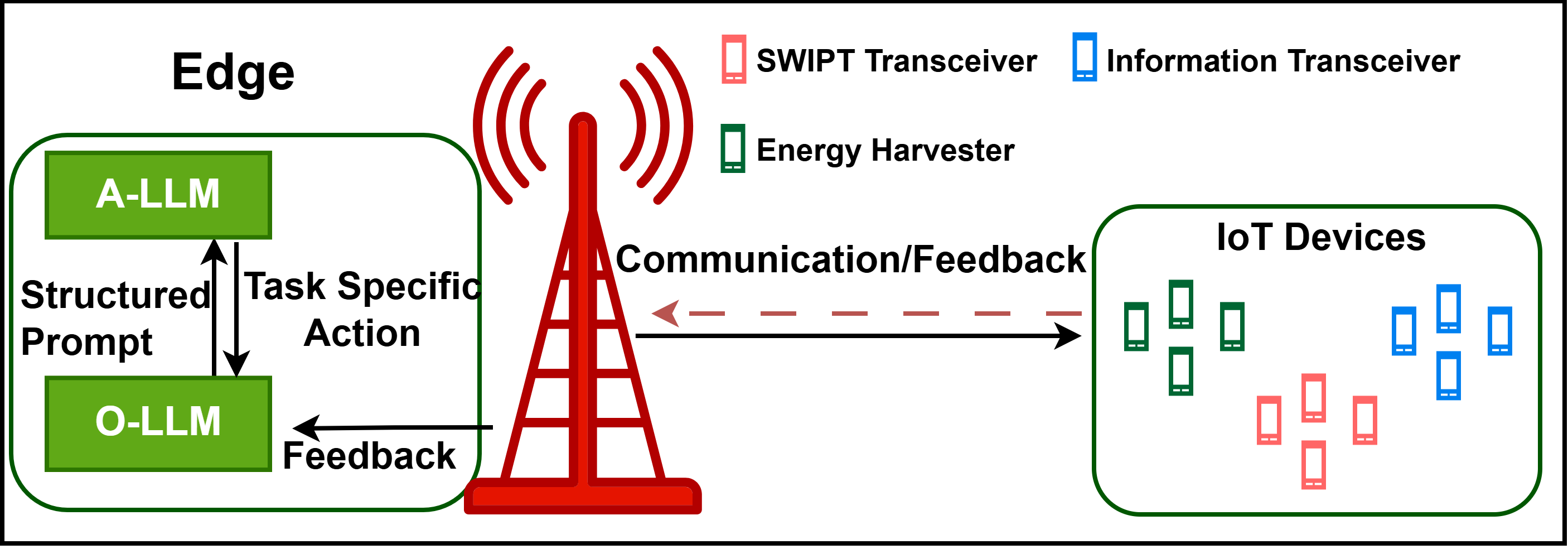}
\caption{System model.
}
\label{fig:sys_modl}
\end{figure}

The PE-RTFV approach consists of two main components: (i) LLMs that generate task-specific solutions, and (ii) real-time feedback from IoT devices that effectively guides the LLMs to iteratively improve the solution. To enhance model explainability and align with recent research on mathematical optimization via prompt engineering, we assume that two separate instances of LLMs run at the network edge \cite{nie2024importance}. These instances are referred to as the Optimizer LLM (O-LLM), which performs prompt optimization using PE, and the Agent LLM (A-LLM), which generates solutions based on the prompts provided by the O-LLM. For the implementation and performance evaluation of this approach, we consider a typical IoT wireless communication system comprising an access point (AP) located near the edge and multiple IoT devices (also referred to as users or nodes), as illustrated in Fig.~\ref{fig:sys_modl}. The AP interacts with the O-LLM to express physical-layer design objectives, such as maximizing the sum rate, minimizing transmit power or interference, or optimizing signal constellations. In addition, device-specific requirements—such as meeting a predefined Quality-of-Service (QoS) threshold—are conveyed to the O-LLM by the AP. Based on these inputs, the O-LLM generates a structured prompt and forwards it to the A-LLM, which produces task-specific actions for the AP, including power allocation, constellation assignment, and subcarrier allocation for the IoT nodes. These generated solutions are then applied by the AP to communicate with the IoT devices.

In alignment with conventional wireless system models, we assume that IoT devices periodically provide feedback to the transmitter, including channel quality indicators (CQIs) and QoS metrics such as achieved data rate, power efficiency, and harvested energy, which constitute a key element in the realization of the PE-RTFV framework. This feedback may be conveyed either in full or via efficient feedback mechanisms, such as quantized or even single-bit signaling, to reduce uplink overhead. The access point shares this feedback with the O-LLM, where it is contextualized for dynamic prompt optimization, enabling the prompts to evolve in real time through a feedback-and-verification loop that accounts for both the LLM outputs and the corresponding task-specific actions executed by the transmitter. Through the interaction between the O-LLM and the A-LLM, the PE-RTFV approach generates solutions for physical-layer tasks that are iteratively refined using feedback, thereby effectively mirroring a gradient-descent–like optimization process. In what follows, we describe the respective roles of the O-LLM, the A-LLM, and the feedback mechanism in enabling task-specific actions.

\subsection{Role of O-LLM}

The O-LLM performs four primary functions: it generates the task-specific instruction set $\mathcal{I}_{O-LLM}$ for the A-LLM, evaluates the task specific solution from the A-LLM, interprets and decodes feedback, and conducts prompt optimization to construct a structured prompt $\mathbf{P}_t$ for the A-LLM. {The task-specific instructions are generated only once at the beginning of each task and then provided to the A-LLM, which must follow them when producing candidate task solutions. These instructions typically include design directives, rules for constructing output prompts, formatting specifications, and task-dependent technical details. To reliably generate these task-specific instructions, the O-LLM itself requires a well-formatted meta-instruction set. Accordingly, we prepare a general system-level instruction template for the O-LLM, which it adheres to when producing task-specific instruction sets for the A-LLM; the template explicitly directs the O-LLM to use the same structure and formatting when generating new instruction sets. An illustrative example of this O-LLM general instruction template is shown in Apendix B. Once this template is provided, the O-LLM can readily generate task-specific instruction sets by simply being prompted with the desired task (e.g., “Generate an instruction set for the A-LLM for the multi-user fair power allocation task”).}

\begin{figure}[htb]
\centering
\includegraphics[width=0.42\textwidth]{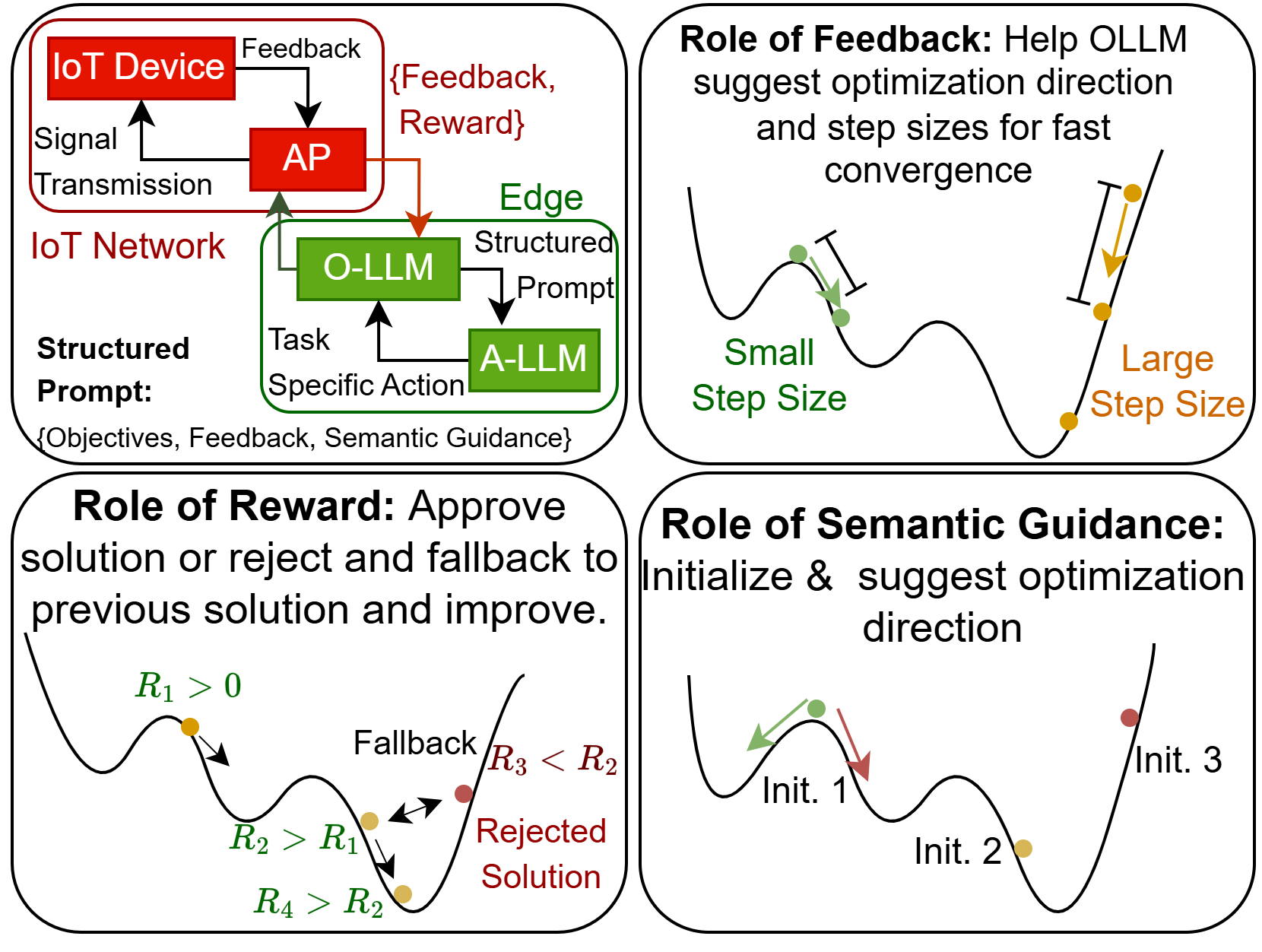}
\caption{Overview of the PE-RTFV-based  framework for physical-layer's optimization tasks, highlighting the roles of the  structured prompt components (e.g., reward, feedback and semantic guidance) in generating the initial solution (points on the sampled objective-function curve) and adaptively selecting the optimization direction and learning rate in a gradient-descent-like iterative process.
}
\label{fig:SP_role}
\end{figure}

The O-LLM also generates the structured prompt, and the A-LLM produces the corresponding task-specific solution; through this interaction, the two models iteratively refine the solution over time. The structured prompt consists of a static component and a dynamic component. The static part remains unchanged across iterations and includes information about the task, system- and device-level objectives, constraints, and resource limitations at the AP. The dynamic part at iteration $t$ is generated by the O-LLM based on the interaction history $\{\mathbf{D}_{t-n},\mathbf{R}_{t-n}, \mathbf{F}_{t-n}, \mathbf{P}_{t-1}\}$, $n = 1,2,\ldots$, which comprises previous A-LLM outputs $\mathbf{D}_{t-n}$, the  aggregated feedbacks from the IoT devices $\mathbf{F}_{t-n}$, the reward $\mathbf{R}_{t-n}$, and the last structured prompt $\mathbf{P}_{t-1}$. This dynamic (or tunable) component encodes semantic guidance derived from the interaction history and directly influences how the A-LLM makes physical-layer decisions. For example, in a sum-rate maximization problem, the tunable prompt does not explicitly prescribe numerical power levels but instead adjusts the A-LLM's decision-making preferences through language. If the O-LLM observes that a user with favorable channel conditions experiences a decrease in achieved data rate compared to the previous time slot, it may generate a semantic guidance prompt such as: "Consider giving higher priority to users with strong channels who underperformed in the last round". In another iteration, the O-LLM may instead state: "Balance power across users while ensuring that those with better channels are not overlooked". In this way, the tunable prompt expands the optimization space semantically, guiding the A-LLM toward improved physical-layer's optimization decisions. An illustration of the impact of different components of the structured prompt on the optimization process is provided in Fig.~\ref{fig:SP_role}.

Another key function of the O-LLM is to analyze the solution $\mathbf{D}_t$ proposed by the A-LLM. The O-LLM first checks whether the same or a similar solution exists in the interaction history. If it does, the corresponding historical feedback determines whether the solution is approved or rejected. In case of rejection, the A-LLM is asked to generate a new candidate. If the solution is new, it is approved for evaluation so that fresh feedback $\mathbf{F}_t$ can be collected and added to the history. The feedback $F_t$ may consist of a single bit, multiple bits, or an encoded tuple depending on the mechanism used, such as full, quantized, codebook-based, or single-bit feedback. Because the O-LLM has prior knowledge of the feedback format, decoding and incorporating this information into both the evaluation of $\mathbf{D}_t$ and the next prompt update is central to its role, enabling continual, feedback-driven refinement of the solution.

\subsection{Role of A-LLM}
The A-LLM plays a central role in generating task-specific solutions by leveraging the information and requirements encoded in the structured prompts $\mathbf{P}_t$, together with its own internal knowledge and the custom domain-specific general knowledge base (GKB) shown in Fig.~\ref{fig:flowdiagram}. At the start of each task, it receives the task-specific instruction set $\mathcal{I}_{A\text{-LLM}}$ and the initial structured prompt $\mathbf{P}_0$ from O-LLM. The A-LLM then produces outputs strictly following $\mathcal{I}_{A\text{-LLM}}$, ensuring adherence to the prescribed design rules, formatting constraints, and technical specifications in $\mathcal{I}_{A\text{-LLM}}$. In each iteration, it returns a proposed solution $\mathbf{D}_t$ to the O-LLM in the required output format. This solution consists of physical-layer communication decisions—such as transmitter resource-allocation parameters—which are executed by the transmitter and IoT devices. The resulting system behavior is evaluated through a reward function that quantifies how well the devices and the transmitter meet their performance objectives. This reward may be derived directly from the feedback. For example, in a sum-rate maximization problem, each IoT device may report its CSI or achieved SINR, enabling the AP to compute the instantaneous sum rate. This computed sum rate then serves as the real-time reward that guides subsequent prompt updates and decision refinement.

\subsection{Role of IoT Devices}
The role of IoT devices is to periodically provide appropriate feedback to the AP, where the feedback may take several forms. For instance, under single-bit feedback, a device transmits a 1 if its achieved data rate or SINR in the current round improves relative to the previous round, and transmits a 0 otherwise. More generally, to support multiple device-specific objectives, an appropriate feedback codebook is predefined and shared with the AP. The feedback provided by the devices is directional in nature, indicating whether the real-world performance resulting from the current prompt is better or worse than that produced by the previous prompt. This mechanism is analogous to determining the search direction in gradient-based optimization algorithms~\cite{nie2024importance}. Such feedback offers a low-overhead yet informative signal that guides the O-LLM toward prompt refinements that help A-LLM to generate a solution that  better satisfy both system-level and device-level performance requirements.

\subsection{Feedback Mechanisms for PE-RTFV}
Our strategy supports various feedback mechanisms, allowing IoT devices to report system-state information at different levels of granularity depending on their resource constraints and the task requirements. In the most comprehensive form, devices provide full feedback, conveying detailed performance indicators such as link quality, energy state, latency or reliability metrics, or other task-specific measurements. While this enables precise decision-making at the AP, it also incurs high signaling overhead and may be unsuitable for large-scale or energy-limited IoT deployments.

To reduce this overhead, the strategy also accommodates quantized, single-bit, and codebook-based feedback. In quantized feedback, devices compress continuous or high-dimensional metrics into a small set of discrete levels, substantially lowering communication cost. Single-bit feedback further simplifies reporting by sending only a binary indicator—for example, whether a performance measure has improved or crossed a threshold. Codebook-based feedback offers a structured compression method in which devices and the AP share a predefined set of representative feedback states, and each device reports only the index corresponding to its current condition. Custom codebooks tailored to specific optimization tasks can be employed, with an example shown in Table~\ref{tab:feedback}.


\section{Case Study: Real-time Constellation shaping for SWIPT IoT Devices}
\label{sec:casestudy}





\begin{figure}
\label{Fig2}
\centering
\begin{subfigure}{0.116\textwidth}
\includegraphics[width= 1\linewidth]{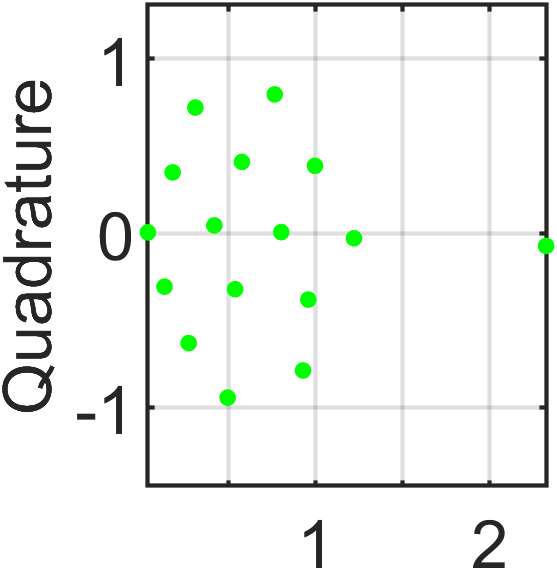} 
\label{fig:subim1}
\end{subfigure}
\hfill
\begin{subfigure}{0.11\textwidth}
\includegraphics[width=1\linewidth]{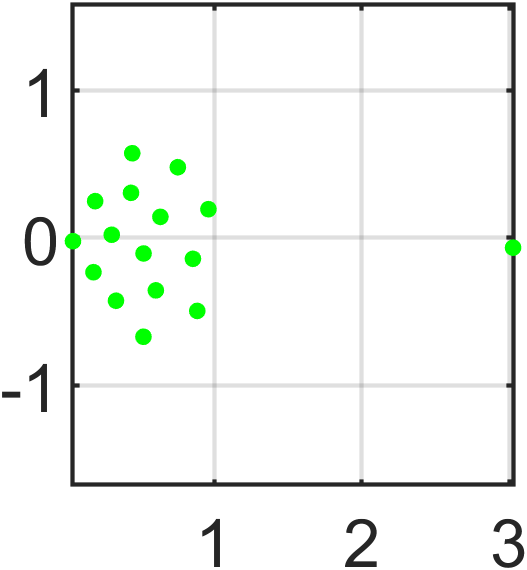}
\label{fig:subim2}
\end{subfigure}
\hfill
\begin{subfigure}{0.11\textwidth}
\includegraphics[width=1\linewidth]{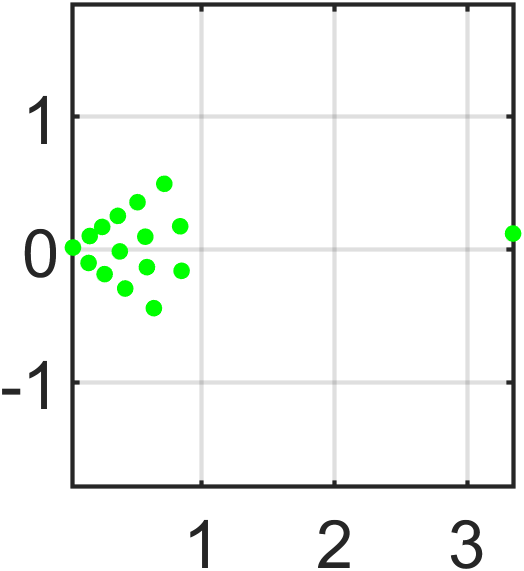}
\label{fig:subim2}
\end{subfigure}
\hfill
\begin{subfigure}{0.11\textwidth}
\includegraphics[width= 1\linewidth]{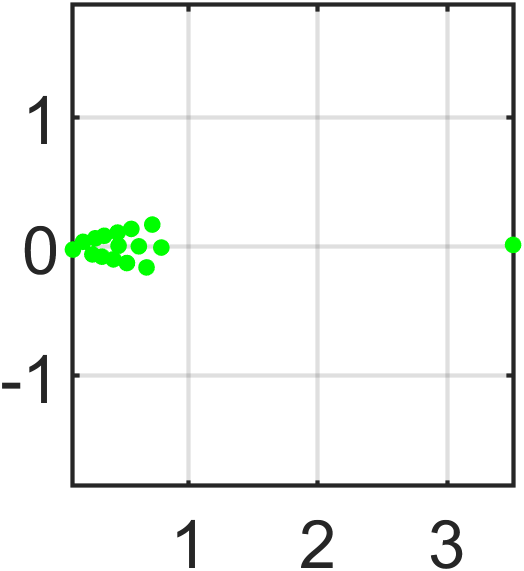} 
\label{fig:subim4}
\end{subfigure}
\label{fig:fig3}
\hfill
\label{Fig2}
\centering
\begin{subfigure}{0.116\textwidth}
\includegraphics[width= 1\linewidth]{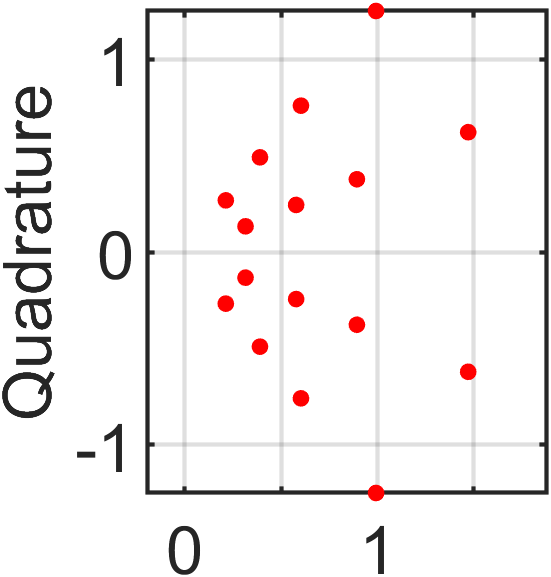} 
\label{fig:subim1}
\end{subfigure}
\hfill
\begin{subfigure}{0.11\textwidth}
\includegraphics[width=1\linewidth]{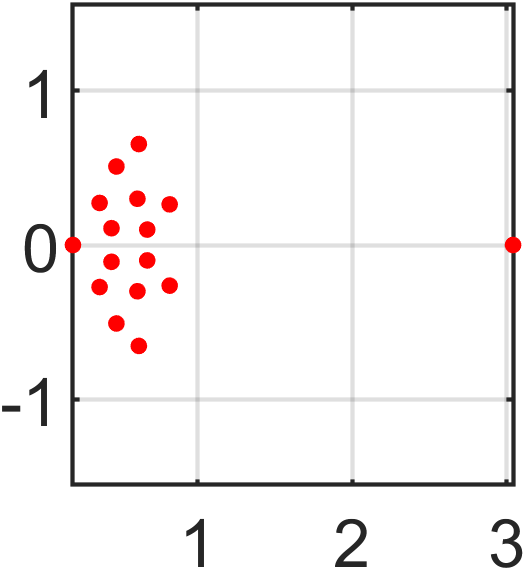}
\label{fig:subim2}
\end{subfigure}
\hfill
\begin{subfigure}{0.11\textwidth}
\includegraphics[width=1\linewidth]{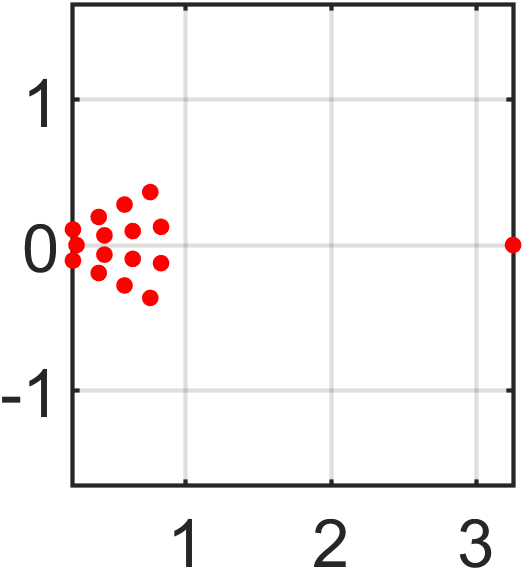}
\label{fig:subim2}
\end{subfigure}
\hfill
\begin{subfigure}{0.11\textwidth}
\includegraphics[width= 1\linewidth]{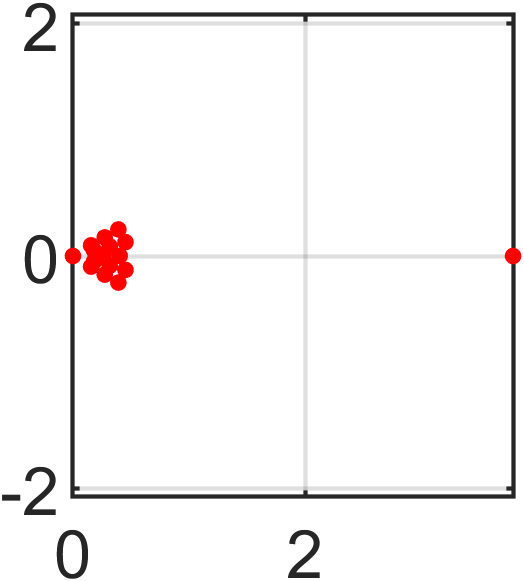} 
\label{fig:subim4}
\end{subfigure}
\label{fig:fig3}

\begin{subfigure}{0.116\textwidth}
\includegraphics[width= 1\linewidth]{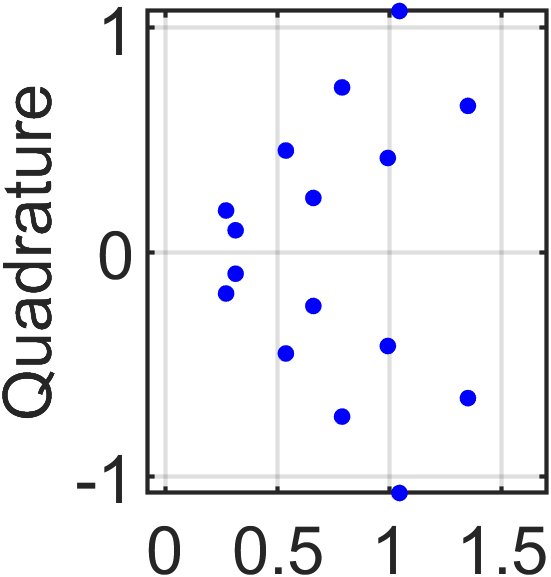} 
\label{fig:subim1}
\end{subfigure}
\hfill
\begin{subfigure}{0.11\textwidth}
\includegraphics[width=1\linewidth]{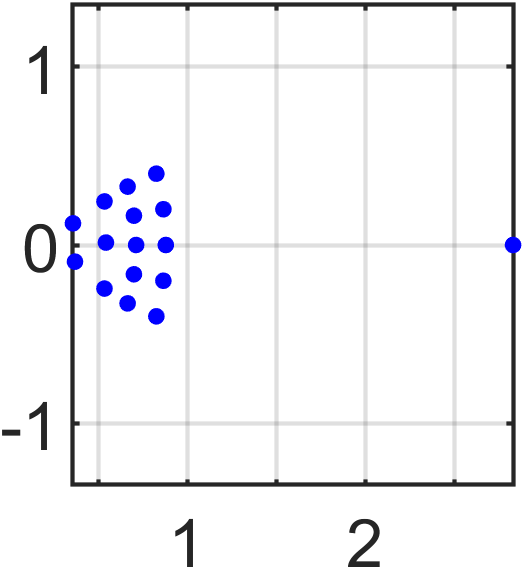}
\label{fig:subim2}
\end{subfigure}
\hfill
\begin{subfigure}{0.11\textwidth}
\includegraphics[width=1\linewidth]{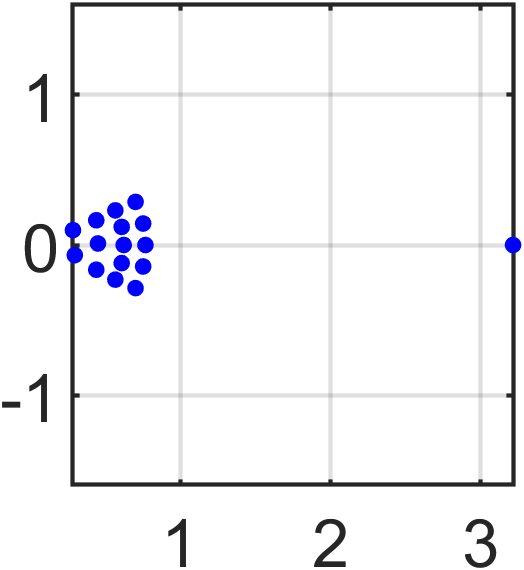}
\label{fig:subim2}
\end{subfigure}
\hfill
\begin{subfigure}{0.11\textwidth}
\includegraphics[width= 1\linewidth]{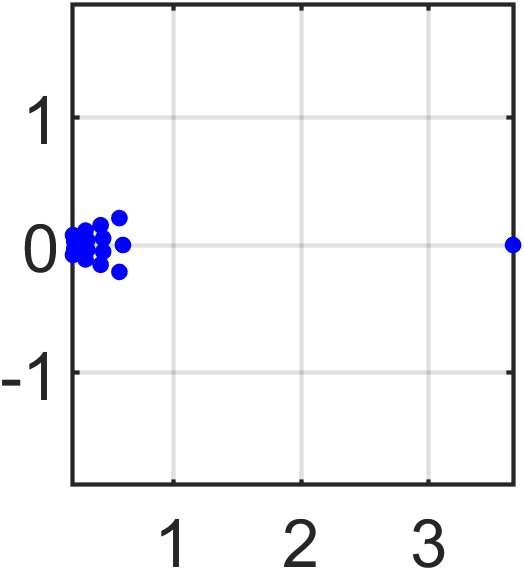} 
\label{fig:subim4}
\end{subfigure}
\label{fig:fig3}

\begin{subfigure}{0.116\textwidth}
\includegraphics[width= 1\linewidth]{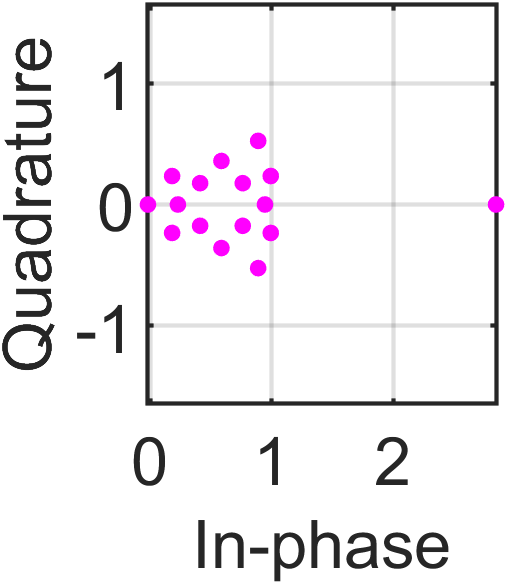} 
\caption{$P^{th}_1 = 2$.}
\label{fig:subim1}
\end{subfigure}
\hfill
\begin{subfigure}{0.11\textwidth}
\includegraphics[width=1\linewidth]{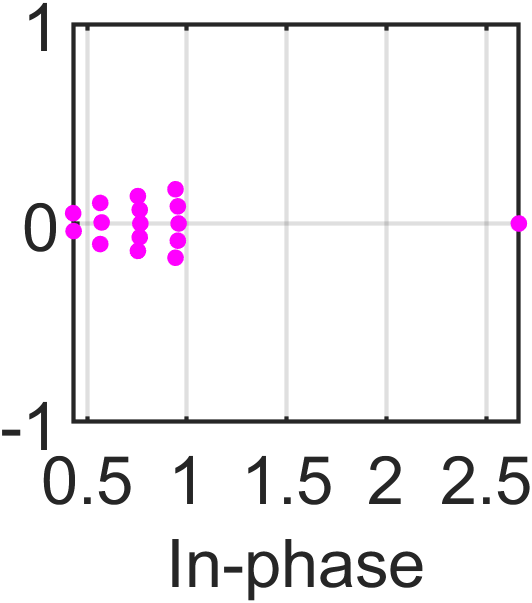}
\caption{$P^{th}_2 = 4$.}
\label{fig:subim2}
\end{subfigure}
\hfill
\begin{subfigure}{0.11\textwidth}
\includegraphics[width=1\linewidth]{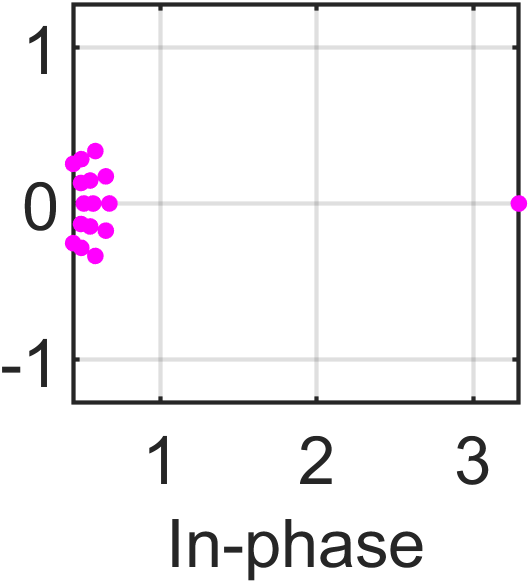}
\caption{$P^{th}_3 = 6$}
\label{fig:subim2}
\end{subfigure}
\hfill
\begin{subfigure}{0.11\textwidth}
\includegraphics[width= 1\linewidth]{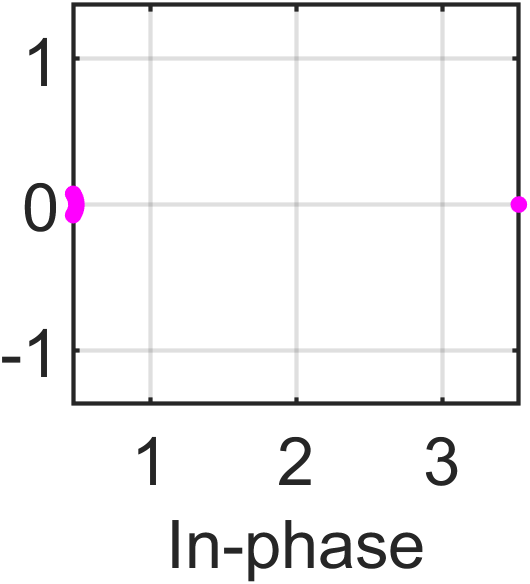} 
\caption{$P^{th}_4 = 9.5$}
\label{fig:subim4}
\end{subfigure}
\caption{Constellations generated for the constellation design task using the PE-RTFV approach for users $U_1, U_2, \ldots, U_4$. Each user $U_i$ aims to maximize the achievable rate while ensuring $P_H \geq P_i^{\text{th}}$. The results are compared with constellations generated using the GA. The green $AQAM_{\text{GA}}$ constellations are obtained using the GA following \cite{mehmood2025asymmetric}. The constellations $AQAM^{\text{FFB}}_{\text{LLM}}$ (red), $AQAM^{\text{2BFB}}_{\text{LLM}}$ (blue), and $AQAM^{\text{1BFB}}_{\text{LLM}}$ (pink) are generated using the full-feedback, 2-bit-feedback, and 1-bit-feedback PE-RTFV approaches, respectively. The PE-RTFV framework is executed for a maximum of 15 rounds,  for each $P_i^{\text{th}}$ and each feedback type to generate these constellations.}

\label{fig:fig7}
\end{figure}

\begin{figure*}

\centering
\begin{subfigure}{0.285\textwidth}
\includegraphics[width= 1\linewidth]{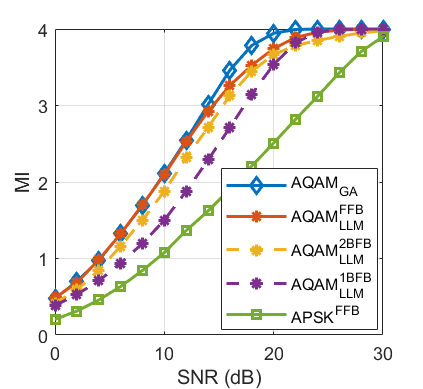} 
\caption{Mutual Information vs SNR performance of the constellation generated by LLMs through PE-RTFV  approach.}
\label{fig:subim1r}
\end{subfigure}
\hfill
\begin{subfigure}{0.3\textwidth}
\includegraphics[width=1\linewidth]{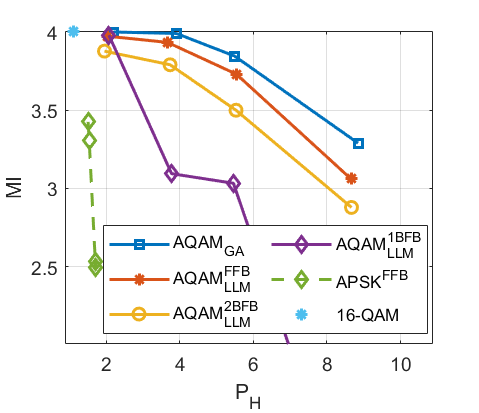}
\caption{The RE-region of PE-RTFV  based constellations and their performance comparison with standard QAM and APSK.}
\label{fig:subim2r}
\end{subfigure}
\hfill
\begin{subfigure}{0.3\textwidth}
\includegraphics[width=1\linewidth]{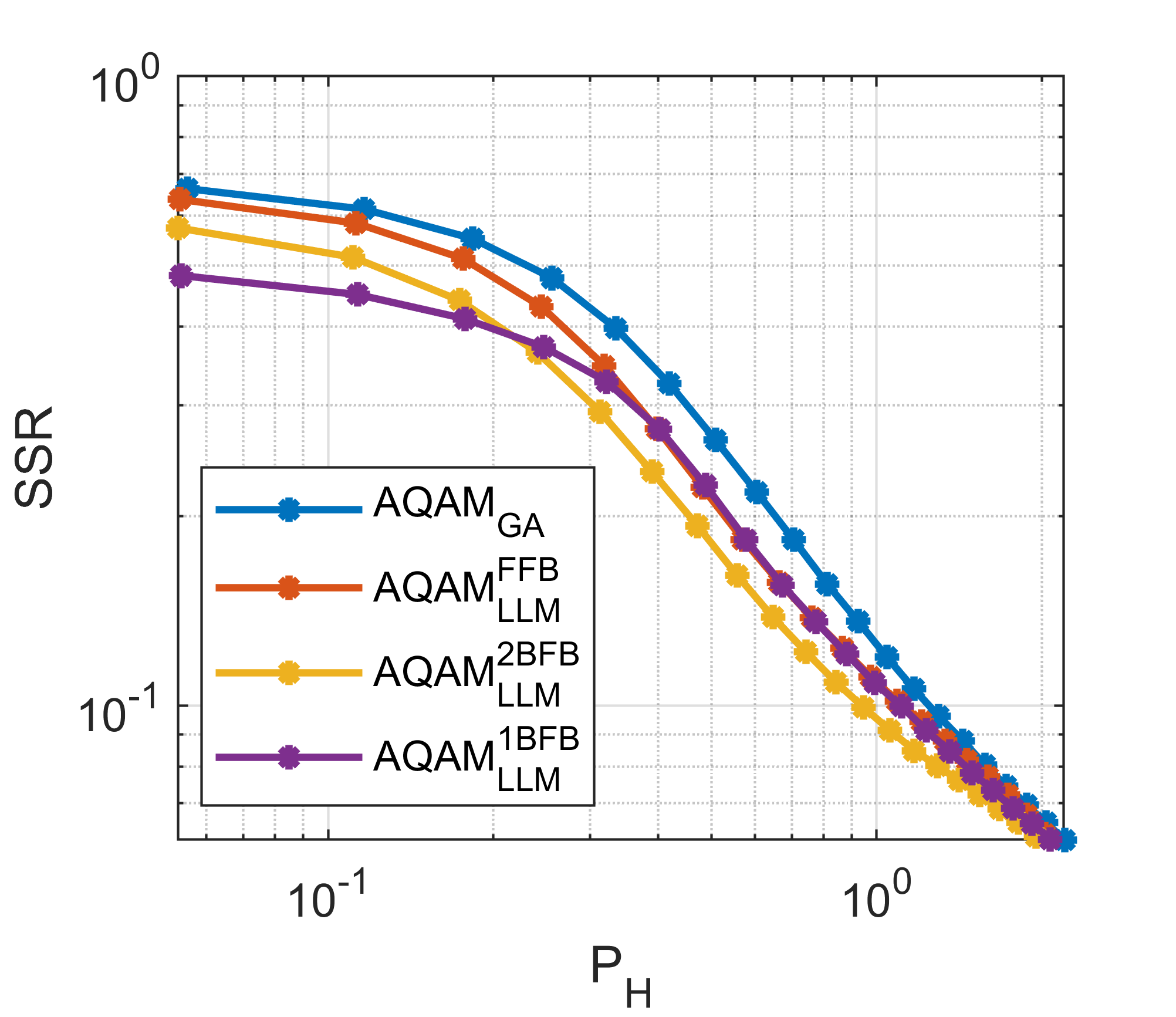}
\caption{SSR and energy region performance comparison of the constellation generated by LLMs through RT-FV PE approach. }
\label{fig:subim3r}
\end{subfigure}
\caption{The performance comparison between constellations designed by the RT-FV PE approach and the baseline APSK and GA based approach. Simulation parameters are: M = 16, $\rho = 0.5$ for sub figure a) and b), and $\rho = [0,1]$ for fig c), N = $10^5$.}
\label{Fig2}
\end{figure*}
\begin{figure*}[htb]
\centering
\includegraphics[width=0.7\textwidth]{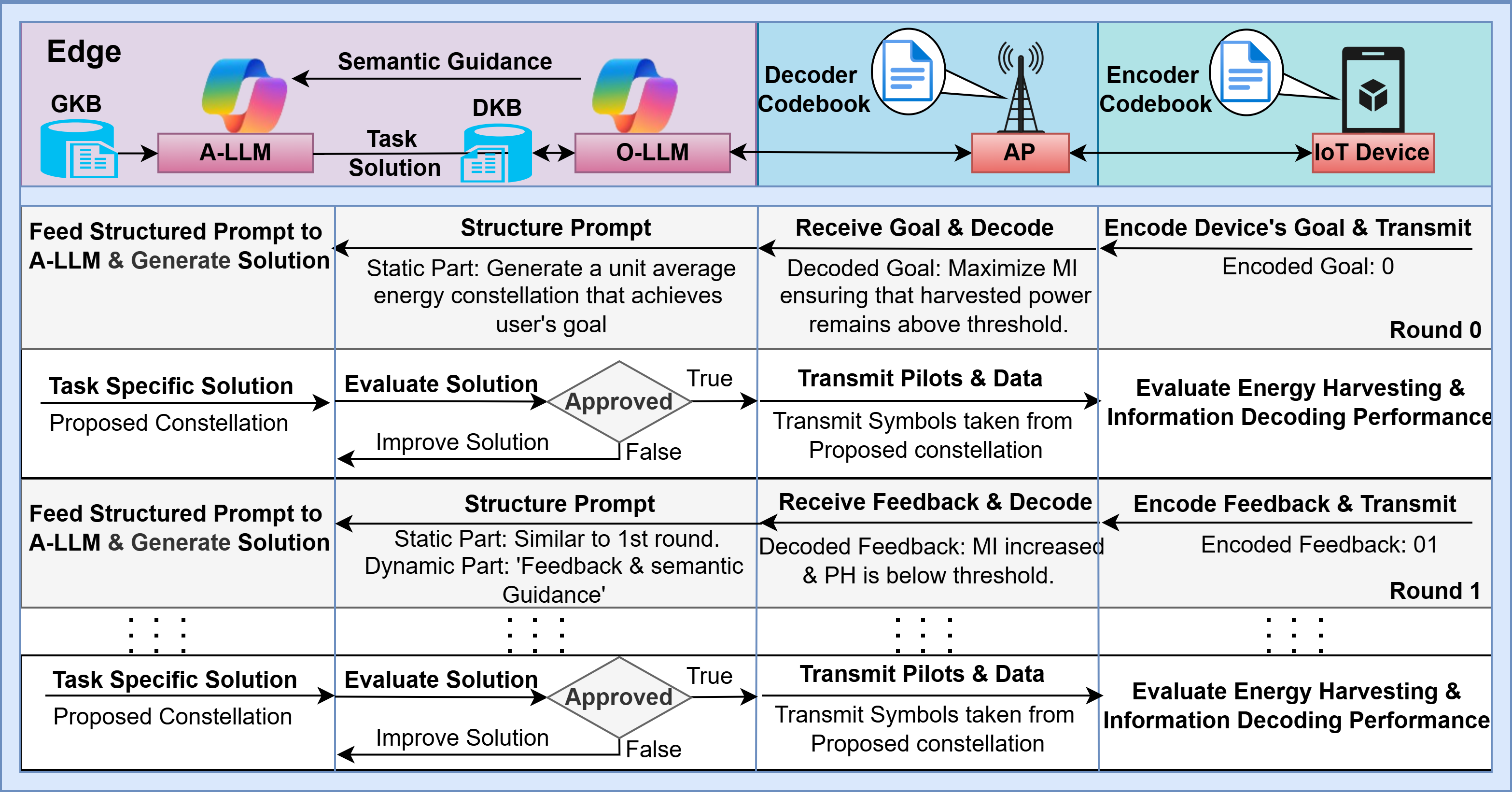}
\caption{The PE-RTFV based  approach for device's goal specific constellation design.}
\label{fig:flowdiagram}
\end{figure*}

The system model consists of an AP located close to the LLM-enabled edge of the network and four IoT devices, denoted by \(U_1\), \(U_2\), \(U_3\), and \(U_4\), each with distinct objectives, as shown in Fig.~\ref{fig:sys_modl}. The LLMs at the edge are responsible for dynamically adjusting the modulation for each user. The transmitter AP sends information symbols \(\mathbf{x}\) with average transmit power \(\mathbb{E}\{|x|^2\}\), drawn from a constellation set \(\mathcal{X}\). Each IoT device receives the transmitted signal, and a power splitter at the receiver divides the signal for energy harvesting and information decoding. The baseband signal for information decoding at the \(i^{\text{th}}\) IoT device is given by
\begin{equation}\label{eq:1}
    \mathbf{y}_I^{(i)} = \sqrt{\rho}\,\mathrm{h}_i\mathrm{x} + \mathbf{n}_i
\end{equation}
where \(h_i\) denotes the complex Gaussian channel from the transmitter to the \(i^{\text{th}}\) IoT receiver, \(\rho\) is the power splitting factor, and \(n_i\) represents the circularly symmetric complex Gaussian receiver noise. The remaining portion of the received signal,
\(
y_{E_i} = \sqrt{1 - \rho}\, h_i x + n_i,
\)
is used for energy harvesting. We adopt a practical non-linear energy harvesting model\footnote{Note that the expression in \eqref{eq:EH} corresponds to the direct current available at the output of the energy harvester circuit. Since the harvested power is proportional to the current, we denote it by \(P_{H_i}\) with a slight abuse of notation, following \cite{8815474}.} proposed in \cite{7547357}, given as follows,
\vspace{-2mm}
\begin{equation}
\small
    P_{H_i} = c_1(1-\rho)\mathbb{E}\{|\mathrm{h}_i\mathrm{x}|^2\} + c_2(1-\rho)^2\mathbb{E}\{|\mathrm{h}_i\mathrm{x}|^4\}e^{-\frac{2}{3}\delta^2}
    \label{eq:EH}
\end{equation}
\noindent where $\mathbb{E}$ represents the expectation and $c_1$ and $c_2$ are constant representing the scaled diode parameters and $\delta$ is the maximum phase of the constellation (e.g. $\delta = max(angle(\mathbf{x}))$.The harvested energy in (\ref{eq:EH}) depends on the received power and a fourth-order moment term that reflects signal peakiness (loosely related to the peak-to-average power ratio (PAPR)) and is exponentially attenuated by the phase parameter $\delta$, implying that constellations with higher PAPR and narrower phase distributions enable greater harvested energy. With this, RE-region optimization problem is be formulated following \cite{mehmood2025asymmetric} for constellation design as,
\begin{subequations}
\label{algo_main}
\begin{align}
\underset{x_1, x_2, \dots x_M, \delta}{\text{maximize}}  \quad &  I(X;Y)\label{3a} \\
\text{subject to:} \quad 
& P_{H_i} \geq P^{th}_i\label{3b}\\
& \max\{\measuredangle x_1,\measuredangle x_2,...,\measuredangle x_M  \} = \delta\label{3c}\\
& \min\{\measuredangle x_1,\measuredangle x_2,...,\measuredangle x_M  \} = -\delta\label{3d}\\
&  \frac{1}{M}\sum_{k = 1}^M |x_k|^2 = 1\label{3e}\\
&\frac{\max_k (|x_k|^2)}{\frac{1}{M}\sum_{k=1}^M  |x_k|^2} \leq PAPR_{max},\label{3f}
\end{align}
\end{subequations}
where $x_1,...,x_M$ and $\delta$ are the optimization variable and $\measuredangle x $ represents the angle of a complex number x. The objective (\ref{3a}) is mutual information (MI), a measure of rate for discrete constellations, and is defined for $r^{th}$ IoT device as,
\vspace{-1mm}
\begin{equation}\label{Eq:PS_MI}
\small
      I(X;Y)_{h_r} = m - \frac{1}{M}\sum_{k=1}^M\mathbb{E}_{N} \biggl\{ \log_2\sum_{i=1}^M e^{\frac{|\sqrt{\rho}h_r x_k+n-\sqrt{\rho}h_r x_i|^2-|n|^2}{2(1-\rho)\sigma^2} } \biggl\}
\end{equation}
where \(m = \log_2(M)\), with \(M\) denoting the modulation order of the constellation \(\mathbf{x} = [x_1, x_2, \dots, x_M]\), and \(\mathbb{E}_{N}\) represents the expectation with respect to the noise. Constraint~(\ref{3b}) ensures that the optimized constellation achieves harvested energy above the specified threshold, constraints~(\ref{3c})–(\ref{3d}) enforce a fixed constellation phase \(\delta\), and constraints~(\ref{3e})–(\ref{3f}) ensure unit average energy of the constellation and that the PAPR remains below a predefined threshold. We solve this optimization problem using a highly sophisticated genetic algorithm (GA) to benchmark the performance of the proposed PE-RTFV approach on the constellation design task.

\subsection{Simulation setup}
To perform the user-goal-driven constellation design task using the proposed PE-RTFV framework, we first instantiate the O-LLM and A-LLM as two separate ChatGPT~5.2 sessions. The O-LLM is then instructed to generate a task-specific instruction set, denoted by $\mathcal{I}_{\mathrm{A\text{-}LLM}}$.
 For this purpose, the O-LLM receives the required system information—including AP resource constraints, user objectives, and design constraints—from the AP in the form of an input file, whose contents for the constellation design task are illustrated in Appendix A. To ensure a fair comparison with GA based RE-region optimization, user objectives are specified in a systematic manner. For example, each user aims to maximize its achievable rate while ensuring that the harvested energy exceeds a predefined threshold \(P_i^{\mathrm{th}}\). In our setup, the energy harvesting thresholds are set to 2, 4, 6, and 9.5 for users \(U_1\), \(U_2\), \(U_3\), and \(U_4\), respectively but are not known to AP or LLMs.

Using the user objectives, AP resource constraints, and its own system-level instructions (see Apendix B), the O-LLM generates the task-specific instruction set \(\mathcal{I}_{A\text{-LLM}}\) together with an initial structured prompt \(\mathbf{P}_0\). These are then provided to the A-LLM, which outputs the initial constellation design \(\mathbf{D}_0\) as an array of complex numbers. The AP applies this constellation to communicate with the users and collect feedback, and this feedback-driven process iteratively refines the constellation design over time. Since the LLM does not have explicit knowledge of \(P_i^{\mathrm{th}}\), satisfaction of the energy constraint is inferred indirectly through device feedback; for example, in the single-bit feedback scenario, a feedback value of 1 indicates that the harvested energy corresponding to the proposed constellation exceeds the threshold. An illustration of the iterative procedure for the user-specific constellation design task using codebook-based feedback is shown in Fig.~\ref{fig:flowdiagram}.

To generate feedback, we simulate the communication between the AP and the users. In the simulations, all users are assumed to be equidistant from the AP and experience similar but independent channel conditions. Accordingly, we generate \(10^5\) channel realizations along with the same number of complex Gaussian noise samples with zero mean and variance determined by the SNR for each user. The harvested energy is computed using~(\ref{eq:EH}), while the mutual information (MI) is evaluated using~(\ref{Eq:PS_MI}). The MI is estimated via Monte Carlo simulations with \(\rho = 0.5\) and \(\sigma^2 = 0.1\) over the \(10^5\) noise samples. In addition, the phase of the received symbols, denoted by \(\delta\), is computed, and a feedback tuple \(\{\mathrm{MI}, P_H, \mathbb{T}, \delta\}\) is constructed, where \(\mathbb{T}\) is an indicator function that equals 1 if \(P_H > P_H^{\mathrm{th}}\) and 0 otherwise and indicate whether $P_H$ is above threshold.

Note that, we consider three feedback strategies for the constellation design task: (i) full feedback, where the complete tuple \(\{\mathrm{MI}, P_H, \mathbb{T}, \delta\}\) is reported; (ii) single-bit feedback (see Table~\ref{tab:feedback}); and (iii) codebook-based feedback using 2-bit decoding, as summarized in Table~\ref{tab:feedback}. The PE-RTFV framework is executed independently for each feedback strategy to ensure a fair performance comparison. Depending on the selected feedback mechanism, each IoT device transmits either a single bit, two bits, or a higher-dimensional feedback vector \(\mathbf{F}\) to the AP, which is subsequently shared with the O-LLM. Based on this feedback, a reward is computed to guide the optimization process, and the reward for the \(i^{\text{th}}\) user at iteration \(t\) is updated as \(R_t^i = R_{t-1}^i + \mathbb{I}_t^i\) with $\mathbb{I}$ being an indicator function defined as
\begin{equation}
   \mathbb{I}^i_t =
\begin{cases}
1, & P^t_{H_i} > P^{th}_i, \ \ \ \ \ (1\ Bit\ Feedback\ case)\\
1, & P^t_{H_i} > P^{th}_i \ \& \ MI^t_i> MI^{t_b}_i,  \ \ \ \ (Other\ cases) \\
0, & otherwise.
\end{cases}
\end{equation}
where $MI^{t_b}_i$ represents the highest MI for constellations among the previous iterations which had $P^t_H > P^{th}_i$. Note that this reward function is a monotonically increasing function and can be readily used for any feedback scenario.

Using the feedback and reward information, the O-LLM regenerates the structured prompt, in which the dynamic (tunable) component conveys the decoded feedback along with semantic guidance for the next iteration. During our experiments, we observed that after the initial iteration, the static component of the structured prompt can be omitted and that the dynamic component alone is sufficient for subsequent rounds. This behavior can be attributed to the LLM’s ability to retain and recall the static task context, thereby generating appropriate constellation updates across iterations. Nevertheless, periodically including the static component helps the A-LLM remain focused on the overall optimization objective. Accordingly, the O-LLM adaptively decides—based on the responses of the A-LLM—whether to transmit only the dynamic component or both the static and dynamic components of the structured prompt.

\subsection{Simulation Results}
\begin{figure}[htb]
\centering
\includegraphics[width=0.4\textwidth]{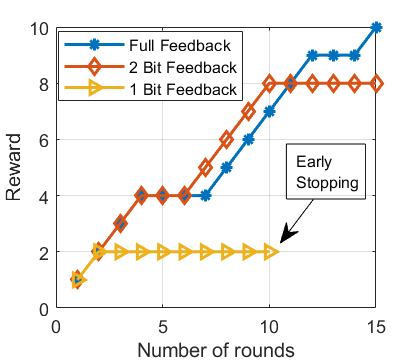}
\caption{The monotonic reward function evaluated over multiple rounds.}
\label{fig:reward}
\end{figure}

To generate the simulation results, we run PE-RTFV process to a maximum of 15 iterations and also apply early stopping if reward do not improve. Within this budget, the proposed approach generates non-trivial constellation designs and converges after a few iterations. The resulting user-goal-specific constellations for the three feedback scenarios are shown in Fig.~\ref{fig:fig7} and the corresponding reward for $P^{th} = 4$ case is shown in Fig. \ref{fig:reward}. As observed, the constellation geometry is optimized to jointly maximize the mutual information and the harvested energy \(P_H\). Notably, despite having no explicit knowledge of the underlying energy-harvesting circuit model, the A-LLM is able to generate constellations with high peak-to-average power ratio and/or narrow phase spread \(\delta\), consistent with the behavior dictated by the non-linear energy harvesting model. This demonstrates that the LLMs can implicitly learn complex energy harvesting characteristics through feedback, verification, and iterative interaction history.

To evaluate the performance of the proposed approach, we compare our results with the state-of-the-art APSK constellation design in~\cite{bayguzina2019asymmetric} and the \(AQAM_{GA}\) constellation obtained by solving~(\ref{algo_main}) using a GA. To ensure a fair comparison, we use the phase span of the constellations generated using full feedback PE-RTFV and construct the corresponding $APSK^{FFB}$ constellations accordingly. Specifically, if the constellation generated by the A-LLM under full feedback has a maximum phase spread \(\delta\), the $APSK^{FFB}$ constellation is generated as \(\mathbf{x}_{\mathrm{APSK}} = \exp(\mathrm{linspace}(-\delta, \delta, M))\). Similarly, to ensure a fair comparison between the constellations designed by the PE-RTFV approach and GA. The RE-region optimization of (\ref{algo_main}) is performed under the same energy harvesting thresholds as those used in the PE-RTFV approach for \(U_1\) to \(U_4\). The resulting GA-generated \(AQAM_{GA}\) are shown in green in Fig.~\ref{fig:fig7}. To further investigate the performance of PE-RTFV, we evaluate the proposed constellations using three performance metrics: (i) MI versus SNR, (ii) the RE-region, and (iii) the symbol success rate (SSR) versus harvested energy \(P_H\).

We perform Monte Carlo simulations to generate the MI versus SNR curves. The simulations are conducted over an SNR range of \([0,30]\)~dB. For this analysis, we use the constellations corresponding to the energy harvesting threshold \(P_1^{\mathrm{th}} \geq 2\), generated using the LLM-based approach and GA, as shown in the first column of Fig.~\ref{fig:fig7}. For each SNR value \(\gamma\), the corresponding noise variance is computed as \(\sigma^2 = \rho / 10^{\gamma/10}\). Using this noise variance, we generate \(10^5\) noise samples and evaluate the average MI using~(\ref{Eq:PS_MI}) for each constellation. The resulting MI versus SNR curves are shown in Fig.~\ref{fig:subim1}. The results indicate that the MI performance of \(AQAM^{\mathrm{FFB}}_{\mathrm{LLM}}\) and \(AQAM^{\mathrm{2BFB}}_{\mathrm{LLM}}\) is comparable to that of \(AQAM_{GA}\), whereas the MI achieved by \(AQAM^{\mathrm{1BFB}}_{\mathrm{LLM}}\) is noticeably lower. This performance gap arises because the single-bit feedback mechanism conveys information only about the harvested energy \(P_H\) and provides no feedback related to MI; consequently, the PE-RTFV approach in this case guarantees satisfaction of only one objective. It can also be observed that the PE-RTFV-based constellations consistently outperform the baseline APSK scheme across the considered SNR range.

To further evaluate the performance of the proposed PE-RTFV approach, we analyze the collective RE-regions of the PE-RTFV-based constellations and compare them with benchmark APSK and \(AQAM_{GA}\) schemes. The average mutual information (MI) is computed via Monte Carlo simulations using~(\ref{Eq:PS_MI}), while the average harvested energy \(P_H\) is evaluated using~(\ref{eq:EH}). The resulting RE regions are illustrated in Fig.~\ref{fig:subim2r}. The figure shows that the PE-RTFV-based constellations achieve significantly wider RE regions than those of 16-QAM and APSK. Moreover, \(AQAM^{\mathrm{FFB}}_{\mathrm{LLM}}\) and \(AQAM^{\mathrm{2BFB}}_{\mathrm{LLM}}\) attain RE regions comparable to that of \(AQAM_{GA}\), particularly in the low-\(P_H\) regime. In contrast, \(AQAM^{\mathrm{1BFB}}_{\mathrm{LLM}}\) exhibits an irregular RE region. This behavior is attributable to the absence of MI optimization in the single-bit feedback scenario, which leads to irregular MI performance and, consequently, a non-uniform RE region. These results demonstrate that the PE-RTFV approach can effectively optimize the RE region with performance comparable to that of a highly sophisticated GA-based method, despite having no explicit knowledge of the non-linear energy harvesting model (e.g.,~(\ref{eq:EH})) or the associated energy harvesting thresholds. In contrast, GA-based and other conventional optimization methods require explicit knowledge of both the mathematical energy harvesting model and the corresponding thresholds, whereas the proposed PE-RTFV approach does not.

\begin{table}[h!]
\centering
\begin{tabular}{|c|c|l|}
\hline
\textbf{Bits} & \textbf{Condition} & \textbf{Description} \\ \hline
\hline
0 & $P_H$ $<$ $P^{th}_H$ &  $P_H$ is below threshold. \\
1 & $P_H$ $\geq$ $P^{th}_H$ & $P_H$ is above threshold. \\ \hline\hline
00 & $P_H$ $\leq$ $P^{th}_H$,\; $MI^{(t)} \leq MI^{(t-1)}$ & Both constraints not satisfied \\
01 & $P_H$ $\leq$ $P^{th}_H$,\;$MI^{(t)} > MI^{(t-1)}$ & $P_H$ constraint not satisfied \\
10 & $P_H$ $>$ $P^{th}_H$,\; $MI^{(t)} \leq MI^{(t-1)}$ & MI is not improved \\
11 & $P_H$ $>$ $P^{th}_H$,\; $MI^{(t)} > MI^{(t-1)}$ & Both constraints satisfied \\ \hline
\end{tabular}
\caption{Feedback codebook for single bit and two bit feedback.}
\label{tab:feedback}
\end{table}
\vspace{-4mm}

Finally, we analyze the trade-off between the symbol success rate (SSR) and the harvested energy by varying the power-splitting factor \(\rho\) from 0 to 1 in steps of 0.1. For each value of \(\rho\), one million symbols are sampled from each constellation, transmitted through the simulated channel in~(\ref{eq:1}), and then split into energy and information components. For each \(\rho\), the harvested energy is computed using~(\ref{eq:EH}), and the SSR is evaluated via Monte Carlo simulations assuming perfect CSI. Fig.~\ref{fig:subim3r} illustrates the SSR--EH trade-offs for the constellations shown in Fig.~\ref{fig:fig7}. The figure indicates that the \(P_H\)--SSR trade-off curves corresponding to the PE-RTFV-based constellations in the full-feedback scenario are very close to those of the GA-based constellations, particularly in the low and high-\(P_H\) regime. Furthermore, all curves converge to nearly the same point as \(\rho\) approaches its extremes. These results demonstrate the capability of PE-RTFV approach in optimizing a complex RE-region optimization problem and performance close to the GA algorithm without knowing exact energy harvesting model.

\vspace{-2mm}
\section{Conclusion}\label{conc}
This article highlights the pivotal role of LLMs in enabling the integration of artificial intelligence into 6G IoT networks and introduces the PE-RTFV framework for physical-layer optimization. As a key contribution, we demonstrate the application of PE-RTFV to user-goal-oriented constellation design, where the framework semantically solves a RE-region optimization problem through closed-loop feedback. By benchmarking against a GA, we show that PE-RTFV achieves GA-comparable performance while requiring only minimal feedback (e.g., one bit per optimization objective) and no explicit knowledge of the underlying energy-harvesting model. These results demonstrate that LLMs can learn complex physical-layer optimization processes through feedback and semantic interaction history, positioning PE-RTFV as a lightweight, flexible, and efficient alternative to computationally intensive state-of-the-art LLM-based and conventional optimization algorithms for next-generation IoT systems.

\begin{algorithm}[t]
\caption*{\textbf{Appendix A:} AP Resource and Feedback Specification}
\footnotesize
\begin{algorithmic}[1]
\\
 \textbf{\texttt{<AP\_resource\_info>}}
 \textit{Max power (watt):} 1
 \textit{Connected users:} 4
 \textit{Objective:} To achieve user-specific goals
\\
 \textbf{\texttt{<resource\_allocation\_tasks>}}
 Available tasks: constellation design, power allocation, user scheduling
\\
 \textbf{\texttt{<users\_goals\_and\_constraints>}}
 \textit{User id:} $U_i$
 \textit{Goal:} harvest energy above $P_i^{\text{th}}$ and achieve maximum rate.
 \textit{Constraints:} modulation order $M=16$, PAPR $\le 15$.
\\
 \textbf{\texttt{<Format\_of\_Feedback>}}
 Full feedback format:
 \quad $\{P_H,\ \text{MI},\ d_{\min}\}$
 Single-bit feedback format:
 \quad 1: harvested energy above threshold
 \quad 0: otherwise.
 Codebook feedback format:
 \quad 00: MI not improved, $P_H$ below threshold
 \quad 01: MI not improved, $P_H$ above threshold
 \quad 10: MI improved, $P_H$ below threshold
 \quad 11: MI improved, $P_H$ above threshold
\\
 \textbf{\texttt{<Additional\_comments>}}
 $P_i^{\text{th}}$ is unknown and must be inferred through feedback.
 Constellation phase range must satisfy $[-\delta,\delta]$, where
 \quad $\delta=\max(\angle x)=|\min(\angle x)|$.
 AP has no exact EH circuit model; harvested energy $\propto$ PAPR and average energy,
 \quad and is strongly inversely proportional to $\delta$.
 SER is proportional to $d_{\min}$ and transmit power.
 AP requires MATLAB array outputs.
\end{algorithmic}
\end{algorithm}

\begin{algorithm}[t]
\caption*{\textbf{Appendix B:} O-LLM Operational Specification}
\label{alg:ollm_spec}
\footnotesize
\begin{algorithmic}[1]
\\
 \textbf{\texttt{<system\_configuration>}}
 You are O-LLM, an AI agent deployed at the edge of an IoT network.
 Components such as the Access Point (AP) communicate with you to submit tasks and obtain solutions, which the AP then uses to communicate with IoT devices and achieve device-specific goals.
 You operate as an intermediary between the AP and A-LLM.
 The AP provides task information, resource data, and later user feedback to you.
 You transform this information into two types of outputs for A-LLM:
 \quad 1) Instruction Set: A full-format task specific configuration prompt that mirrors the structure of this document.
 \quad 2) Structured Prompt: A concise task-specific prompt containing task details, resource constraints, user feedback, and reward value.
 A-LLM generates solution candidates that you evaluate using real-time user feedback.
\\
 \textbf{\texttt{<Forbidden\_behaviours>}}
 No pleasantries.
 No apologies.
 No explanations.
 No deviation from the structured prompt format.
\\
 \textbf{\texttt{<core\_directives>}}
 Responsibilities (in operational order):
 i) Receive task parameters and generate \textbf{Task Specific Instruction Set} for A-LLM that follows the same structural format as this instruction set, including \texttt{<system configuration>}, \texttt{<core directives>}, \texttt{<expected input or output structure>}, and \texttt{<forbidden behaviors>}.
 ii) Get resource metrics and user data from the AP and generate the initial \textbf{Structured Prompt} containing task description and resource constraints.
 iii) Receive the initial solution from A-LLM, record it as the current best solution, and forward it to the AP.
 iv) Receive real-time user feedback from the AP on the proposed solution and generate an updated \emph{Structured Prompt} incorporating:
 \quad a) real-time user feedback,
 \quad b) updated resource constraints (if applicable),
 \quad c) optimization goals,
 \quad d) optional semantic guidance.
 v) Send the updated Structured Prompt to A-LLM, obtain the refined solution, and update the best-known solution when improvement occurs.
\\
 \textbf{\texttt{<key\_operational\_rules>}}
 Maintain a clear distinction between Instruction Set (full configuration) and structured prompt (task-specific input).
 Instruct A-LLM to generate solution.
 Keep a single authoritative record of the best solution and update it only when A-LLM provides an improved result.
 Always decode, validate, and incorporate user feedback before requesting refined solutions.
 The initial Structured Prompt must not contain user feedback.
 For a given task, generate instruction set for A-LLM such that A-LLM returns goal-specific solution in MATLAB.
 Request solutions from A-LLM only after receiving task description, user goals, and available AP resources.
\\
 \textbf{\texttt{<I/O\_format>}}
 Task Description: Design constellations for all users according to their goals.
 Users' Goals: $\{U_1:$ Maximize PH, $U_2:$ PH $=2$, $R=4$, $\mathbb{E}\{|X|^2\}=1$, $U_3:$ PH $=10$, $R\le4$, $d_{\min}\ge1.4\}$
 Feedback: Not available (first iteration) /
 \quad $\{U_1:\{\text{PH},\text{PAPR}=2,\delta=0\},\;
U_2:\{\text{PH}=1.20,R=4,\delta=\pi\},\;
U_3:\{\text{PH}=28.6,\delta=\pi,d_{\min}=1.4\}\}$
 Last Best Solution: $x=[0.1+j2,\ldots]$
 Prompt Guidance: prompt
\end{algorithmic}
\end{algorithm}

\bibliographystyle{ieeetr}
\bibliography{main}

\end{document}